\documentclass[epj]{svjour}
\usepackage{graphics}
\usepackage{epsfig}
\newfont{\cmsy}{cmsy12}
\begin{document}
\title{Determination of the Dalitz plot parameter $\alpha$ for the decay $\eta \rightarrow 3\pi^0$ with the Crystal Ball at MAMI-B}
%\subtitle{Do you have a subtitle?\\ If so, write it here}
\author{M.~Unverzagt\inst{1}\fnmsep\inst{2}\fnmsep\thanks{\emph{e-mail}: unvemarc@kph.uni-mainz.de},
	P.~Aguar-Bartolom\'{e}\inst{1},
	J.~Ahrens\inst{1},
	J.R.M.~Annand\inst{3},
	H.J.~Arends\inst{1},
	R.~Beck\inst{1}\fnmsep\inst{2},
	V.~Bekrenev\inst{4},
	B.~Boillat\inst{5},
	A.~Braghieri\inst{6},
	D.~Branford\inst{7},
	W.J.~Briscoe\inst{8},
	J.W.~Brudvik\inst{9},
	S.~Cherepnya\inst{10},
	R.~Codling\inst{3},
	E.J.~Downie\inst{1}\fnmsep\inst{3},
	L.V.~Fil'kov\inst{10}
	D.I.~Glazier\inst{7},
	R.~Gregor\inst{11},
	E.~Heid\inst{1},
	D.~Hornidge\inst{12},
	O.~Jahn\inst{1},
	V.L.~Kashevarov\inst{10},
	R.~Kondratiev\inst{13},
	M.~Korolija\inst{14},
	M.~Kotulla\inst{11},
	D.~Krambrich\inst{1},
	B.~Krusche\inst{5},
	M.~Lang\inst{1}\fnmsep\inst{2},
	V.~Lisin\inst{13},
	K.~Livingston\inst{3},
	S.~Lugert\inst{11},
	I.J.D.~MacGregor\inst{3},
	D.M.~Manley\inst{15},
	M.~Martinez-Fabregate\inst{1},
	J.C.~McGeorge\inst{3},
	D.~Mekterovic\inst{14},
	V.~Metag\inst{11},
	B.M.K.~Nefkens\inst{9},
	A.~Nikolaev\inst{1}\fnmsep\inst{2},
	R.~Novotny\inst{11},
	R.O.~Owens\inst{3},
	P.~Pedroni\inst{6},
	A.~Polonski\inst{13},
	S.N.~Prakhov\inst{9},
	J.W.~Price\inst{9},
	G.~Rosner\inst{3},
	M.~Rost\inst{1},
	T.~Rostomyan\inst{6},
	S.~Schumann\inst{1}\fnmsep\inst{2},
	D.~Sober\inst{16},
	A.~Starostin\inst{9},
	I.~Supek\inst{14},
	C.M.~Tarbert\inst{7},
	A.~Thomas\inst{1},
	Th.~Walcher\inst{1},
	D.P.~Watts\inst{7},
	F.~Zehr\inst{5}
	\\ \\
	(Crystal Ball at MAMI, TAPS and A2 Collaborations)
}                     % Do not remove

\titlerunning{Dalitz plot parameter $\alpha$ for the decay $\eta \rightarrow 3\pi^0$}
\authorrunning{M.~Unverzagt \textit{et al.}}

\institute{Institut f\"ur Kernphysik, University Mainz, Mainz, Germany
	\and Helmholtz-Institut f\"ur Strahlen- und Kernphysik, University Bonn, Bonn, Germany
	\and Department of Physics and Astronomy, University of Glasgow, Glasgow, United Kingdom
	\and Petersburg Nuclear Physics Institute, Gatchina, Russia
	\and Institut f\"ur Physik, University Basel, Basel, Switzerland
	\and INFN Sezione di Pavia, Pavia, Italy
	\and School of Physics, University of Edinburgh, Edinburgh, United Kingdom
	\and Center for Nuclear Studies, The George Washington University, Washington, D.C., USA
	\and University of California at Los Angeles, Los Angeles, California, USA
	\and Lebedev Physical Institute, Moscow, Russia
	\and II. Physikalisches Institut, University Gie\ss en, Gie\ss en, Germany
	\and Mount Allison University, Sackville, NB, Canada
	\and Institute for Nuclear Research, Moscow, Russia
	\and Rudjer Boskovic Institute, Zagreb, Croatia
	\and Kent State University, Kent, Ohio, USA
	\and The Catholic University of America, Washington, D.C., USA
}

\date{Received: date / Revised version: date}
% The correct dates will be entered by Springer
%
\abstract{
	A precise measurement of the Dalitz plot parameter, $\alpha$, for the $\eta \rightarrow 3\pi^0$ decay is presented. The experiment was performed with the Crystal Ball and TAPS large acceptance photon detectors at the tagged photon beam facility of the MAMI-B electron accelerator in Mainz. High statistics of $1.8 \cdot 10^6$ $\eta \rightarrow 3\pi^0$ events were obtained, giving the result $\alpha = -0.032 \pm 0.002_{\mathrm{stat}} \pm 0.002_{\mathrm{syst}}$.
 % end of PACS codes
} %end of abstract
\maketitle

\section{Introduction}\label{hdintro}

The $\eta \rightarrow 3\pi^0$ decay violates isospin symmetry. Therefore, it offers a unique possibility to study symmetries and symmetry-breaking characteristics of strong interactions. Because electromagnetic contributions to the amplitude can be neglected \cite{Sut66,Bau96,Dit08}, this decay occurs due to the isospin breaking part of the QCD Hamiltonian:
\begin{equation}
        \mbox{\cmsy H \usefont{T1}{ptm}{m}{n}}_{\not{\:\mathrm{I}}} = \frac{1}{2}(m_{\mathrm{u}}-m_{\mathrm{d}})(\bar{\mathrm{u}}\mathrm{u} - \bar{\mathrm{d}}\mathrm{d}).
	\label{eqIsoBreak}
\end{equation}
Therefore, the amplitude is proportional to the mass difference $m_\mathrm{u} - m_\mathrm{d}$ of the two lightest quarks u and d.

Calculations of the decay amplitude are usually based on the framework of Chiral Perturbation Theory ($\chi$PT). The leading order term $ \mbox{\cmsy O \usefont{T1}{ptm}{m}{n}}(p^2)$ of the momentum expansion yields the constant amplitude \cite{Osb70}:
\begin{equation}
	A(\eta \rightarrow 3\pi^0) = \frac{B_0 (m_\mathrm{u}-m_\mathrm{d})}{\sqrt{3}F_{\pi}^2} \sim (m_\mathrm{u}-m_\mathrm{d}),
	\label{eqAmpp2}
\end{equation}
where $B_0$ and the pion decay constant $F_\pi$ are the low-energy constants of $\chi$PT. There is further theo\-retical work determining the amplitude in the second \cite{Gas85} and the third \cite{Bij07} chiral order, both of which include loop diagrams describing final-state rescatterings of the pions. Higher order rescattering effects are examined by using dispersion methods \cite{Kam96,Ani96} or the unitarised $\chi$PT approach (UCHPT) \cite{Bor05} based on the Bethe-Salpeter equation.

The squared absolute value of the decay amplitude may be expanded around the centre of the Dalitz plot \cite{PDG08} and Bose symmetry dictates the form: 
\begin{equation}
	|A(\eta \rightarrow 3\pi^0)|^2 = |N|^2 (1 + 2 \alpha z + \ldots).
	\label{eqAmpExpand}
\end{equation}
$N$ is a normalisation constant, which is equal to the amplitude that would apply, if the decay proceeded only according to the available phase space. The Dalitz plot parameter, $\alpha$, describes the pion energy dependence of the squared absolute value of the decay amplitude up to first order of the expansion. The parameter $z$ is given by \cite{PDG08}
\begin{equation}
	z = 6 \sum_{i=1}^3 \left( \frac{E_i-m_{\eta}/3}{m_{\eta}-3m_{\pi^0}}\right)^2 = \frac{\rho^2}{\rho_{max}^2}.
	\label{eqz}
\end{equation}
Here $E_i$ represents the pion energies in the $\eta$ rest frame and $\rho$ is the distance from the centre to a point in the Dalitz plot. $\rho_{max}$ is the maximum value of $\rho$. $z$ varies from $z=0$, where all three pions have the same energy $E_i = m_{\eta}/3$, to $z=1$, where one of the pions is at rest. So, determining the Dalitz plot parameter offers a nice possibility to test the different theoretical calculations based on $\chi$PT (see table \ref{tbtheo}).

\begin{table}
	\caption{Theoretical results for the Dalitz plot parameter, $\alpha$.}
	\label{tbtheo}       % Give a unique label
	\begin{center}
	\begin{tabular}{lcc}
		\hline\noalign{\smallskip}
		Calculation & Refs. & $\alpha$ \\
		\noalign{\smallskip}\hline\noalign{\smallskip}
		$\chi$PT $ \mbox{\cmsy O \usefont{T1}{ptm}{m}{n}}(p^2)$ & \cite{Bij02} & 0  \\
		$\chi$PT $ \mbox{\cmsy O \usefont{T1}{ptm}{m}{n}}(p^4)$ & \cite{Bij02} & $0.015$ \\
		$\chi$PT $ \mbox{\cmsy O \usefont{T1}{ptm}{m}{n}}(p^6)$ & \cite{Bij07} & $0.013 \pm 0.032$ \\
		Dispersion & \cite{Kam96} & $-0.007 \ldots -0.014$ \\
		UCHPT & \cite{Bor05} & $-0.031 \pm 0.003$ \\
	\noalign{\smallskip}\hline
	\end{tabular}
	\end{center}
\end{table}

In lowest order of $\chi$PT there is no final-state interaction between the three pions. Thus, the amplitude is constant \cite{Osb70} and $\alpha = 0$ \cite{Bij02}. With the amplitude including one-loop diagrams \cite{Gas85} Bijnens \cite{Bij02} calculated $\alpha = 0.015$. Including two-loop contributions, Bijnens and Ghorbani \cite{Bij07} get $\alpha = 0.013 \pm 0.032$. Contrary to the experimental results summarised in table \ref{tbexp}, all calculations based on $\chi$PT alone predict a positive value. Using instead dispersion relations \cite{Kam96} in combination with extended Khuri-Treiman equations \cite{Khu60} gives the result $-0.014 \leq \alpha \leq -0.007$, depending on the input parameters of the calculation. The unitarised $\chi$PT approach \cite{Bor05} based on the Bethe-Salpeter equation constrains its free parameters to known branching ratios and $\pi \pi$ scattering phases from \cite{Eid04}. The fit yields $\alpha = -0.031 \pm 0.003$, within one standard deviation in agreement with the high-statistics experimental results of the Crystal Ball at BNL \cite{Tip01} and KLOE collaborations \cite{Amb07}.

Table \ref{tbexp} summarises the experimental world data set for the Dalitz plot parameter, $\alpha$. The Crystal Ball collaboration \cite{Tip01} measured $\alpha = -0.031 \pm 0.004$ from a sample of $9.5 \cdot 10^5$ events obtained at BNL, while KLOE \cite{Amb07} found $\alpha = -0.027 \pm 0.004^{+0.004}_{-0.006}$ with $6.5 \cdot 10^5$ events. All other experiments listed in table \ref{tbexp} collected not more than $1.2 \cdot 10^5$ $\eta \rightarrow 3\pi^0$ events. The PDG \cite{PDG08} currently quotes $\alpha = -0.031 \pm 0.004$, but this value is dominated by the result of the Crystal Ball collaboration, since the latest publication of the KLOE collaboration has not been included.

\begin{table}
	\caption{Experimental results for the Dalitz plot parameter, $\alpha$. The result of Baglin \textit{et al.} \cite{Bag70} has been omitted due to a lack of statistical significance.}
	\label{tbexp}       % Give a unique label
	\begin{center}
	\begin{tabular}{lcc}
		\hline\noalign{\smallskip}
		Experiment & Refs. & $\alpha$ \\
		\noalign{\smallskip}\hline\noalign{\smallskip}
		Crystal Ball at BNL & \cite{Tip01} & $-0.031 \pm 0.004$ \\
		KLOE & \cite{Amb07} & $-0.027 \pm 0.004^{+0.004}_{-0.006}$ \\
		GAMS-2000 & \cite{Ald84} & $-0.022 \pm 0.023$ \\
		Crystal Barrel & \cite{Abe98} & $-0.052 \pm 0.017 \pm 0.010$ \\
		SND & \cite{Ach01} & $-0.010 \pm 0.021 \pm 0.010$ \\
		CELSIUS/WASA & \cite{Bas07} & $-0.026 \pm 0.010 \pm 0.010$ \\
		WASA at COSY & \cite{Ado08} & $-0.027 \pm 0.008 \pm 0.005$ \\
	\noalign{\smallskip}\hline
	\end{tabular}
	\end{center}
\end{table}

When proposing the measurement of the $\eta \rightarrow 3\pi^0$ decay at MAMI, only the result of one high-statistics experiment, the Crystal Ball at BNL, was published. This lack of precise data led to the conclusion that a new high-statistics measurement of the Dalitz plot parameter was needed, again using the Crystal Ball, but with the $\eta$-photoproduction mechanism instead of a hadronic pion beam at BNL. This need for another precise measurement of the Dalitz plot parameter was emphasised even more, when the KLOE collaboration announced their preliminary result $\alpha = -0.014 \pm 0.004_{\mathrm{stat}} \pm 0.005_{\mathrm{syst}}$ \cite{Cap05}, which showed a big discrepancy with the BNL value. The KLOE collaboration at DA$\phi$NE used $\mathrm{e}^+ \mathrm{e}^- \rightarrow \phi \rightarrow \eta \gamma$ as the production reaction. Although the revised KLOE result \cite{Amb07} agrees with the Crystal Ball at BNL value within the given errors, the fundamental importance of the Dalitz plot parameter still requires further experimental input.

This paper describes an analysis of tagged photon experiments carried out at the electron accelerator MAMI-B (Mainz Microtron B) \cite{Her83,Wal90} in the years 2004 and 2005 within the Crystal Ball/TAPS collaboration \cite{Unv08}. Data from two different experiments with the CB/TAPS-setup, that differed in the trigger conditions and the tagged photon energy range, were analysed. The first was dedicated to the neutral decays of the $\eta$ meson and especially to the rare $\eta$ decays. The second investigated radiative $\pi^0$-photo\-production.

\section{Experimental setup}\label{hdsetup}

The Dalitz plot parameter, $\alpha$, was determined from $3\pi^0$ decays of $\eta$ mesons produced in the $\gamma \mathrm{p} \rightarrow \eta \mathrm{p}$ reaction. The photons were emitted by bremsstrahlung of 883\,MeV electrons from the MAMI-B \cite{Her83,Wal90} accelerator. The electrons were separated from the photons and momentum analysed by the Glasgow tagged photon spectrometer \cite{Ant91,Hal96} at Mainz, with an energy resolution of approximately 2\,MeV. The photon energies were determined by energy conservation with a maximum photon flux above the $\eta$ threshold of roughly $1 \cdot 10^5 \gamma/(\mathrm{s\;MeV})$.

\begin{figure}
        \begin{center}
                \includegraphics[scale=0.4, clip]{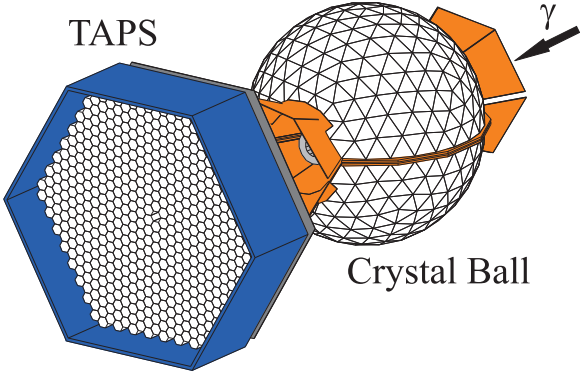}
                \caption{The Crystal Ball/TAPS setup.}
                \label{psSim}
        \end{center}
\end{figure}

The 4.8\,cm long liquid hydrogen target was located at the centre of the spherical Crystal Ball (CB) \cite{Ore82,Sta01} photon spectrometer. The CB consisted of 672 optically insulated NaI(Tl) crystals, each read out by an individual photomultiplier. It covered the full azimuthal angle range for polar angles between 20\,$^\circ$ and 160\,$^\circ$. Each crystal had the shape of a 41\,cm (15.7 radiation lengths) long truncated pyramid pointing towards the centre of the CB. Electromagnetic showers were measured with an energy resolution of $\sigma/E_\gamma = 0.02/(E_\gamma /\mathrm{GeV})^{1/4}$. The resolution in the polar angle $\sigma_\theta$ was $2^\circ$ to $3^\circ$, while for the azimuthal angle it was $\sigma_\phi = \sigma_\theta / \sin \theta$. The target at the centre of the CB was surrounded by a thin particle identification detector (PID) \cite{Wat04} to register charged particles hitting the CB. It consisted of 24 30\,cm long and 2\,mm thick optically isolated plastic scintillator bars arranged parallel to the beam axis, so that each covered $15^\circ$ of the azimuthal angle range.

The forward angles between $\theta = 4^\circ$ and $\theta = 20^\circ$ were covered by TAPS \cite{Nov91,Gab94} consisting of 510 BaF$_2$ crystals arranged as a wall, preceded by a single layer of 510 5\,mm thick veto plastic scintillators. The CB and TAPS together covered roughly 97\,\% of the total solid angle (fig. \ref{psSim}). TAPS was positioned 173\,cm downstream of the centre of the CB, giving the opportunity for an efficient time-of-flight analysis for particle identification. Each of the hexagonally shaped BaF$_2$ crystals had an inner diameter of 5.9\,cm and a length of 25\,cm, which corresponds to approximately 12 radiation lengths. The shower energy resolution was $\sigma /E_\gamma = 0.0079/(E_\gamma /\mathrm{GeV})^{0.5}+0.018$. The angular resolution for 300\,MeV photons was $0.7^\circ$ full width at half maximum (FWHM).

The experiment trigger comprised two event conditions. First, the total sum of the CB photomultiplier analogue signals had to exceed a threshold that corresponded to approximately 390\,MeV for one experiment and 60\,MeV for the other. Secondly, the sector multiplicity in the CB and TAPS had to be greater than 2. Here up to 16 adjacent crystals made up a sector in the CB, giving 45 hardwired sectors. In TAPS each sector was one quarter of the wall. If at least one of the signals in the sector exceeded a threshold of 20 to 40\,MeV, depending on the relative calibration of the photomultiplier signals, it contributed to the multiplicity.

The determination of the Dalitz plot parameter requires the elimination of the phase space contribution to the amplitude. Therefore, a Monte Carlo simulation of the experiment was produced with $\alpha = 0$. For the analy\-sis described in this paper $100 \cdot 10^6$ $\eta \rightarrow 3\pi^0$ events were generated and the particle propagation simulated with the GEANT v3.21 software. These events were analysed in the same way as the experimental data, resulting in an average reconstruction efficiency of $\varepsilon_{\eta \rightarrow 3\pi^0} \approx 23\,\%$.

\section{Data analysis}\label{hdanalysis}

The Dalitz plot parameter, $\alpha$, for the decay $\eta \rightarrow 3\pi^0$ was calculated from the analysis of the reaction
\begin{equation}
	\gamma \mathrm{p} \rightarrow \eta \mathrm{p} \rightarrow 3\pi^0 \mathrm{p} \rightarrow 6\gamma \mathrm{p}.
\end{equation}
The first step of the event selection demanded six clusters in the CB and TAPS identified as photons, ignoring the other particle types like protons, charged pions and neutrons. With the CB and the PID, clusters were identified as charged particles by checking the agreement of the azimuthal angles of CB clusters with the $\phi$ angle of the hit PID elements. Photons had no corresponding PID hit. The charged particles were then divided into protons and charged pions by comparing the energy of the clusters with the energy deposit in the PID scintillator. Neutrons could not be separated from photons in the CB. In TAPS photons, protons, charged pions and neutrons could be identified using information from the veto detectors, comparing the time of flight with the energy deposition in the BaF$_2$ crystals and analysing the pulse-shape of the BaF$_2$ signals. A cluster was formed by a group of adjacent crystals, which had registered parts of the electromagnetic shower initiated by an incoming particle. But a crystal could only contribute to a cluster, if its energy deposit exceeded 2\,MeV in the CB and 4\,MeV in TAPS. The energy of the cluster was calculated by the sum of the energies of all contributing crystals. The cluster direction was determined by the weighted sum of the directions of the contributing crystals, using the square-root of the energies as weight. Only clusters, which had a total energy of 20\,MeV or higher, were used in the analysis.

\begin{figure*}
        \begin{center}
                \includegraphics[scale=0.4]{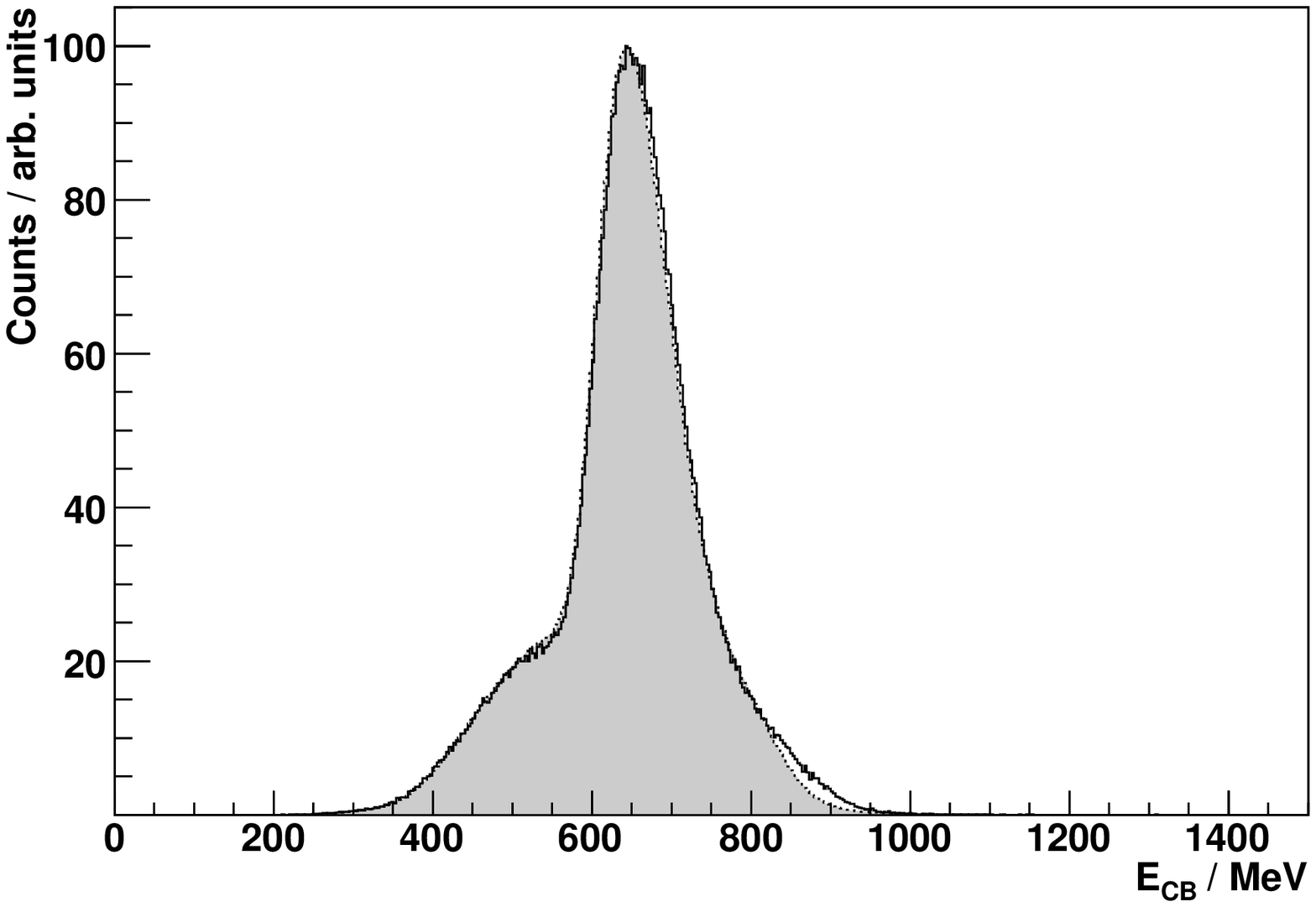}\includegraphics[scale=0.4]{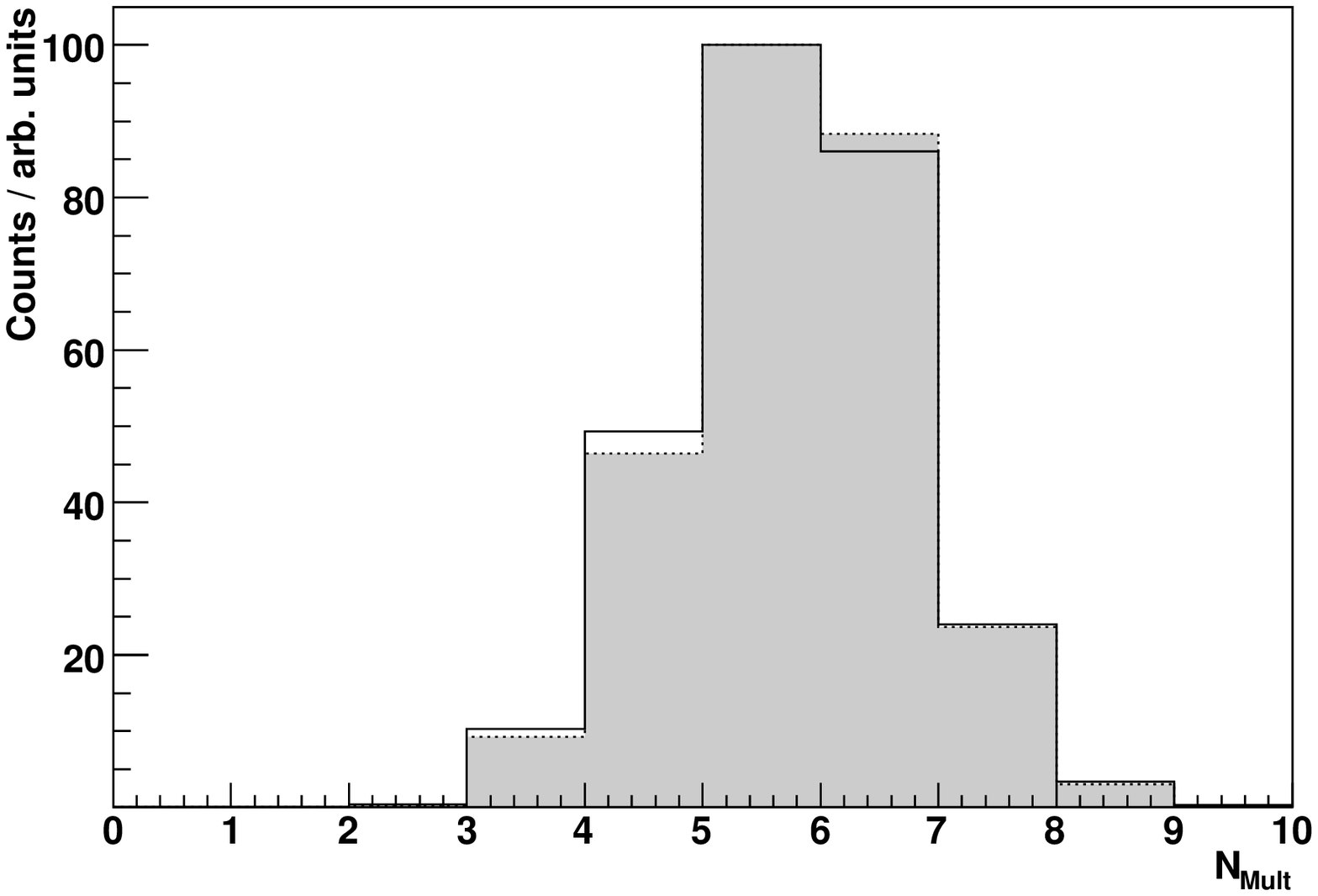}
                \caption{The two trigger conditions. The grey shaded histograms were calculated from the simulated data. The black line corresponds to the experiment. \textit{Left}: CB energy sum. \textit{Right}: Sector multiplicity.}
                \label{psTrig}
        \end{center}
\end{figure*}

\begin{figure*}
	\begin{center}
		\includegraphics[scale=0.4]{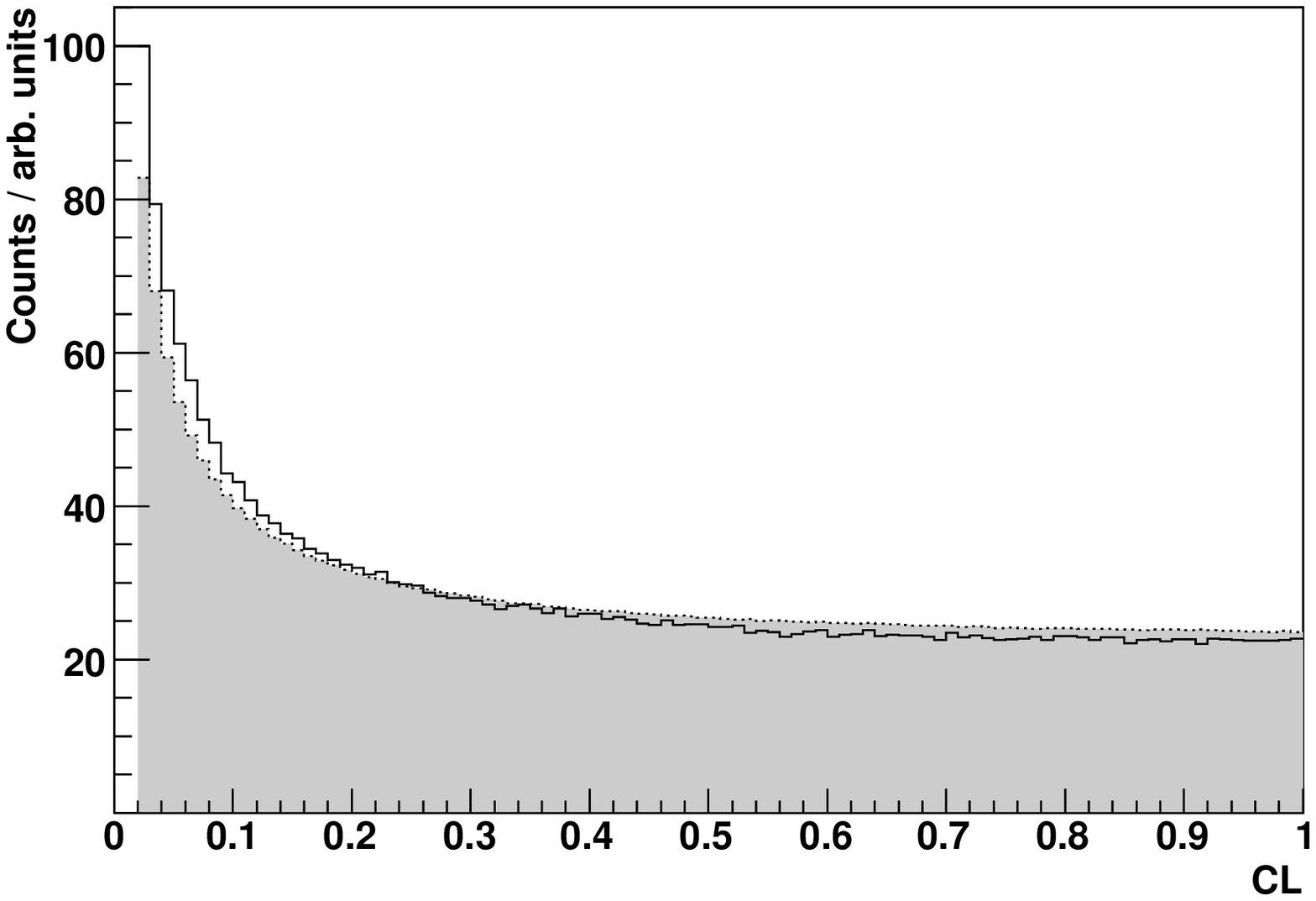}\includegraphics[scale=0.4]{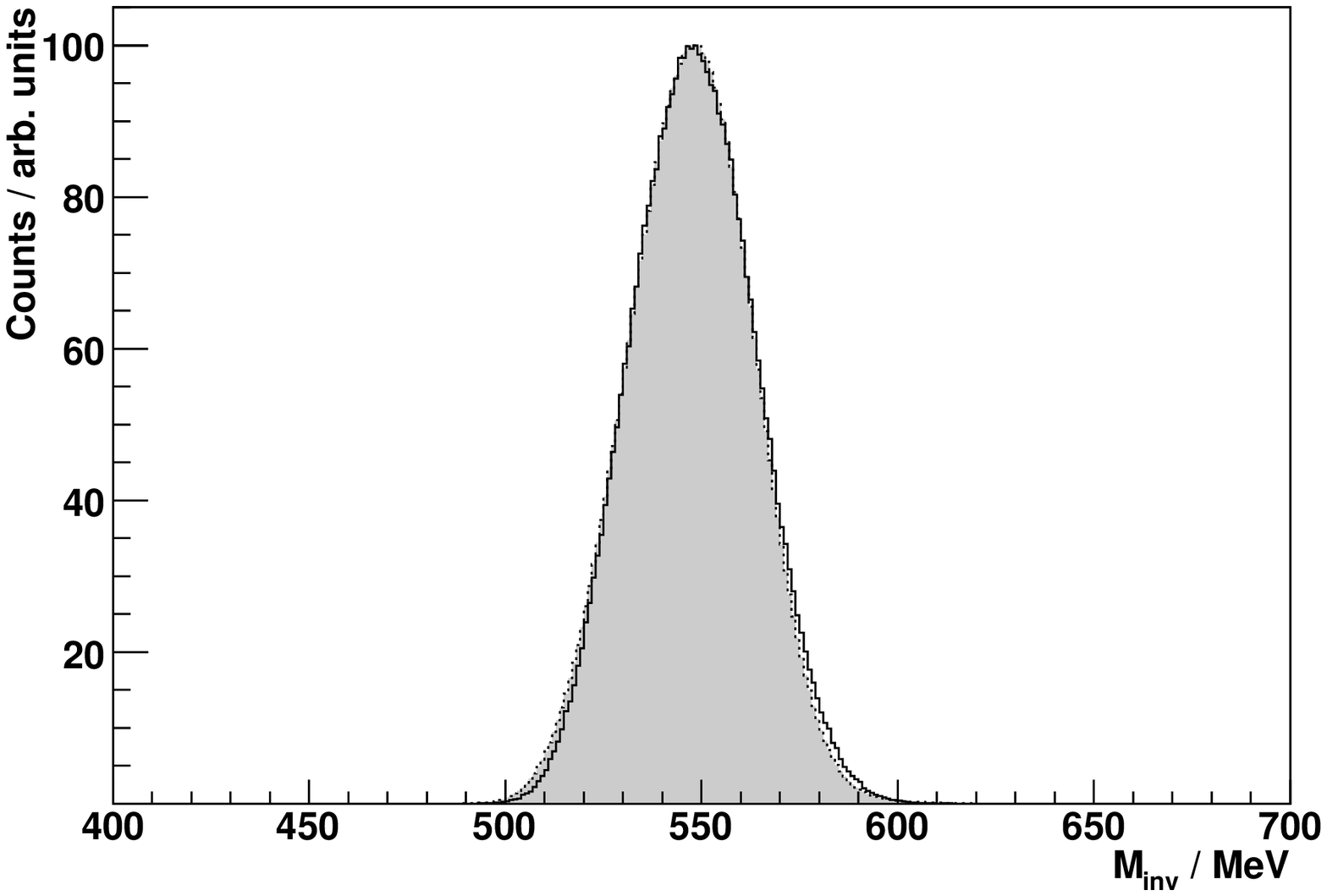}
		\includegraphics[scale=0.4]{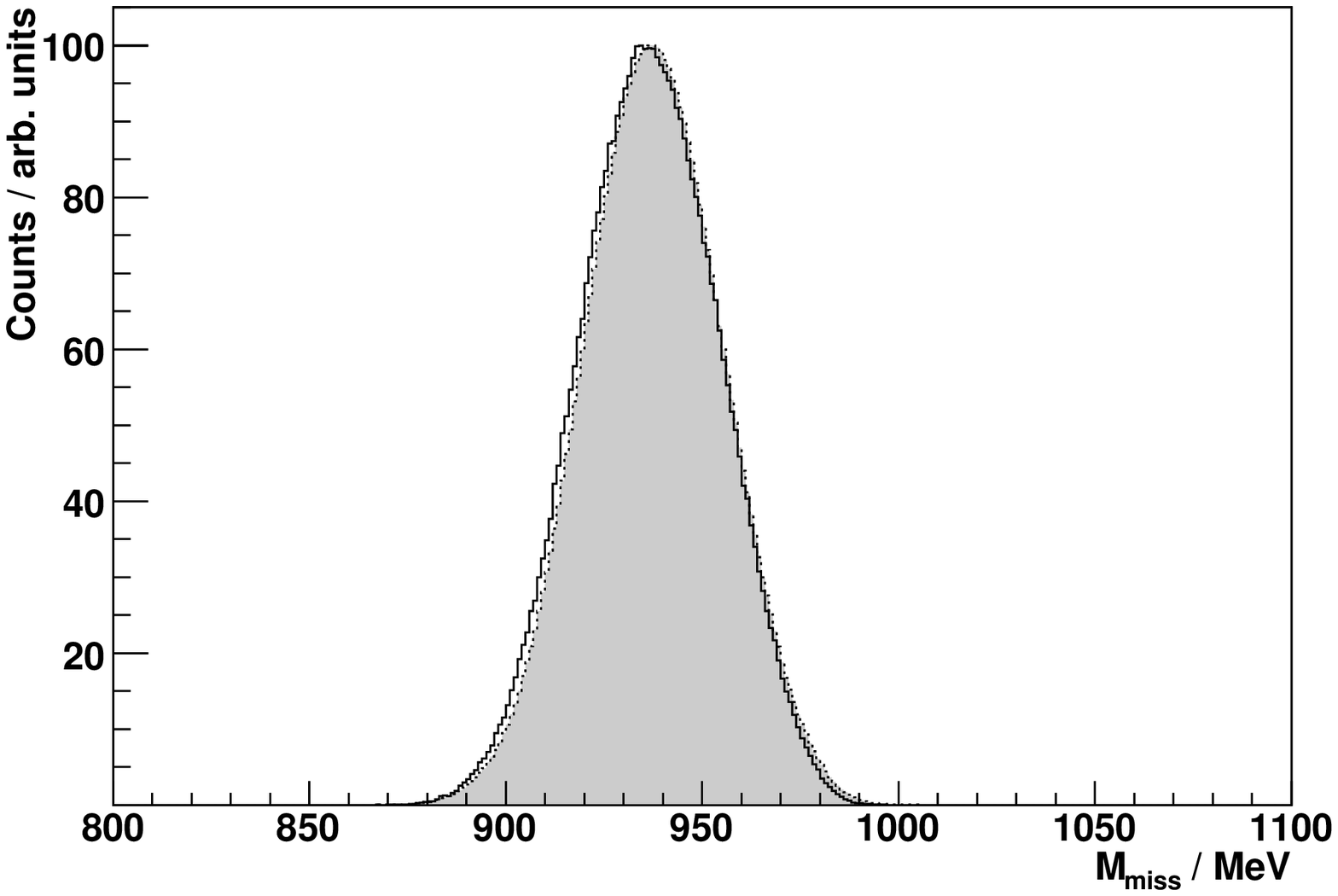}\includegraphics[scale=0.4]{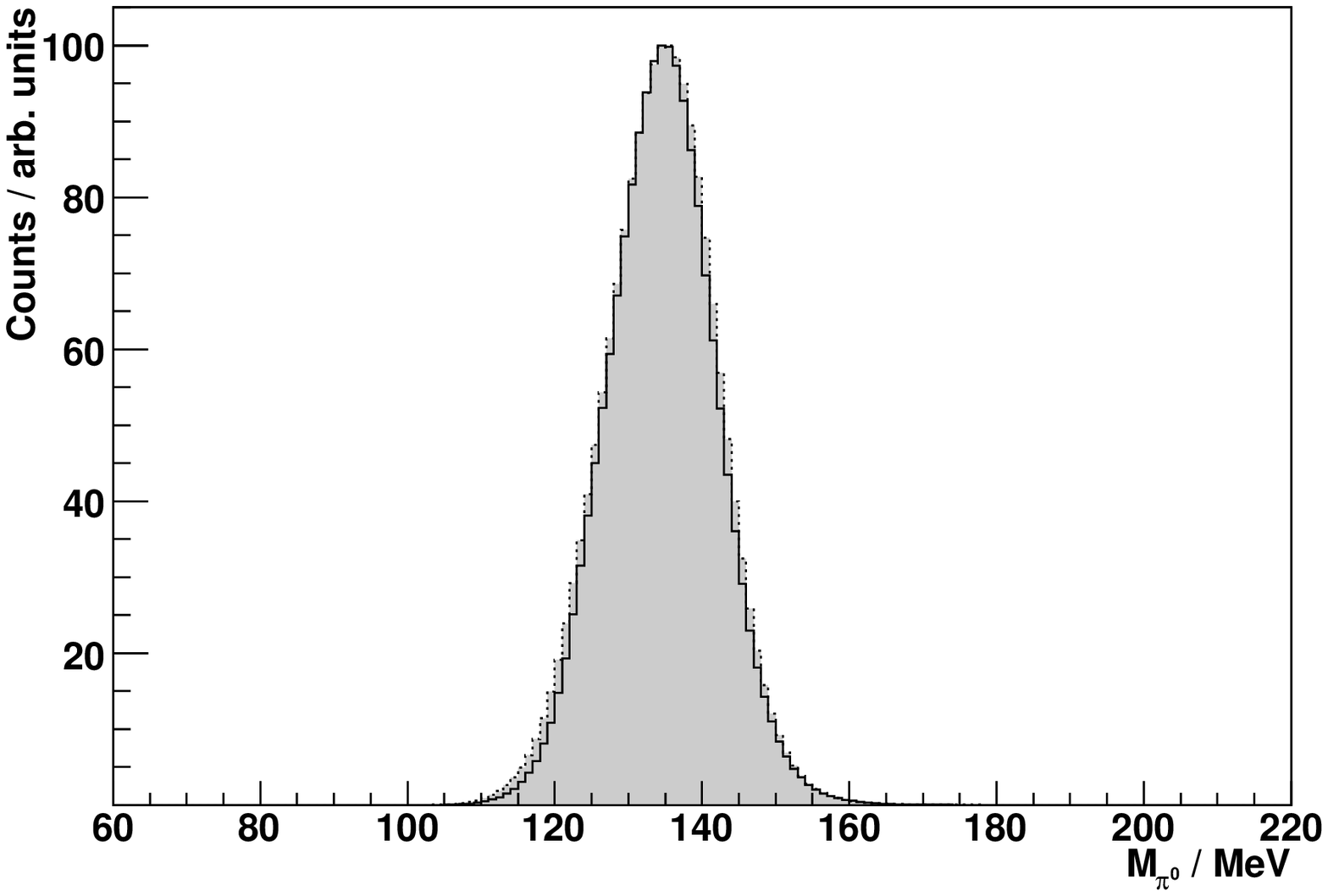}
		\caption{Kinematic fit results of the selected events. The grey shaded histograms and the solid lines represent the simulation and the experimental data, respectively. All masses were obtained from the directly measured energies and momenta (not using the fitted quantities). \textit{Top left}: Confidence level. \textit{Top right}: Invariant mass of the six photons. \textit{Bottom left}: Missing mass of the six photons. \textit{Bottom right}: Invariant masses of the three photon pairs.}
		\label{psFit}
	\end{center}
\end{figure*}

The next step in the analysis of the simulated data was to implement a software trigger system. This was crucial, since electronics and photomultipliers of the PID in the exit region of the CB screened TAPS from particles emerging from the target. Some photons were converted into electron positron pairs in this material and subsequently identified as charged particles in TAPS. Protons lost parts of their energy in this inactive material. But these effects could be simulated precisely as discussed in \cite{Unv08}. Figure \ref{psTrig} shows the agreement between the experimental and simulated trigger conditions for selected $\eta \rightarrow 3\pi^0$ events. Note, that a sector in the trigger did not directly correspond to a cluster in the calorimeters. Trigger sectors had a much tighter restriction due to the high thresholds on single crystals compared to the total energy threshold on the clusters. Therefore, more than half of the selected events had a multiplicity lower than six.

In further analysis of the experimental and the simulated data, the reaction hypothesis $\eta \rightarrow 3\pi^0 \rightarrow 6\gamma$ was tested with a kinematic fitting technique \cite{Ave91}. The measured energies $E$ and the $\theta$ and $\phi$ angles were used as input parameters to a fit with five constraints: the invariant and the missing masses of the six identified photons had to be equal to the masses of the $\eta$ meson and the proton, respectively, and the invariant masses of each of the three photon pairs had to give the $\pi^0$ mass. All 15 possible combinations to form three pairs from six photons were tested in separate fits. Events in which at least one fit had a confidence level (CL) higher than 2\,\% were considered as $\eta \rightarrow 3\pi^0$ events. This cut was chosen to reject most of the background, but lose only a small fraction of events of interest. The adjusted photon energies and angles from the fit with the highest CL were used to calculate the Dalitz plot parameter, $\alpha$.

The resolutions of the three variables $E$, $\theta$ and $\phi$ used in the fits were determined by a Monte Carlo simulation of the CB/TAPS-setup. Photons of different energies were generated emerging isotropically from the centre of the CB and the detector response was simulated in GEANT. The resolutions $\sigma_E$, $\sigma_\theta$ and $\sigma_\phi$ were then obtained by comparing the reconstructed variables with the initially generated values.

Since the events were selected by cutting on the CL of the fits, it was important that the simulated CL distribution agreed with the measured distribution. Then cuts made at different confidence levels removed the same fraction of events from the measured and the simulated samples. Figure \ref{psFit} shows good agreement between the two distributions (top left). The rise at low confidence levels was produced by events where parts of the electromagnetic shower leaked out of the detector, such as when clusters were located at the edge of the CB.

\begin{figure*}
	\begin{center}
		\includegraphics[scale=0.4]{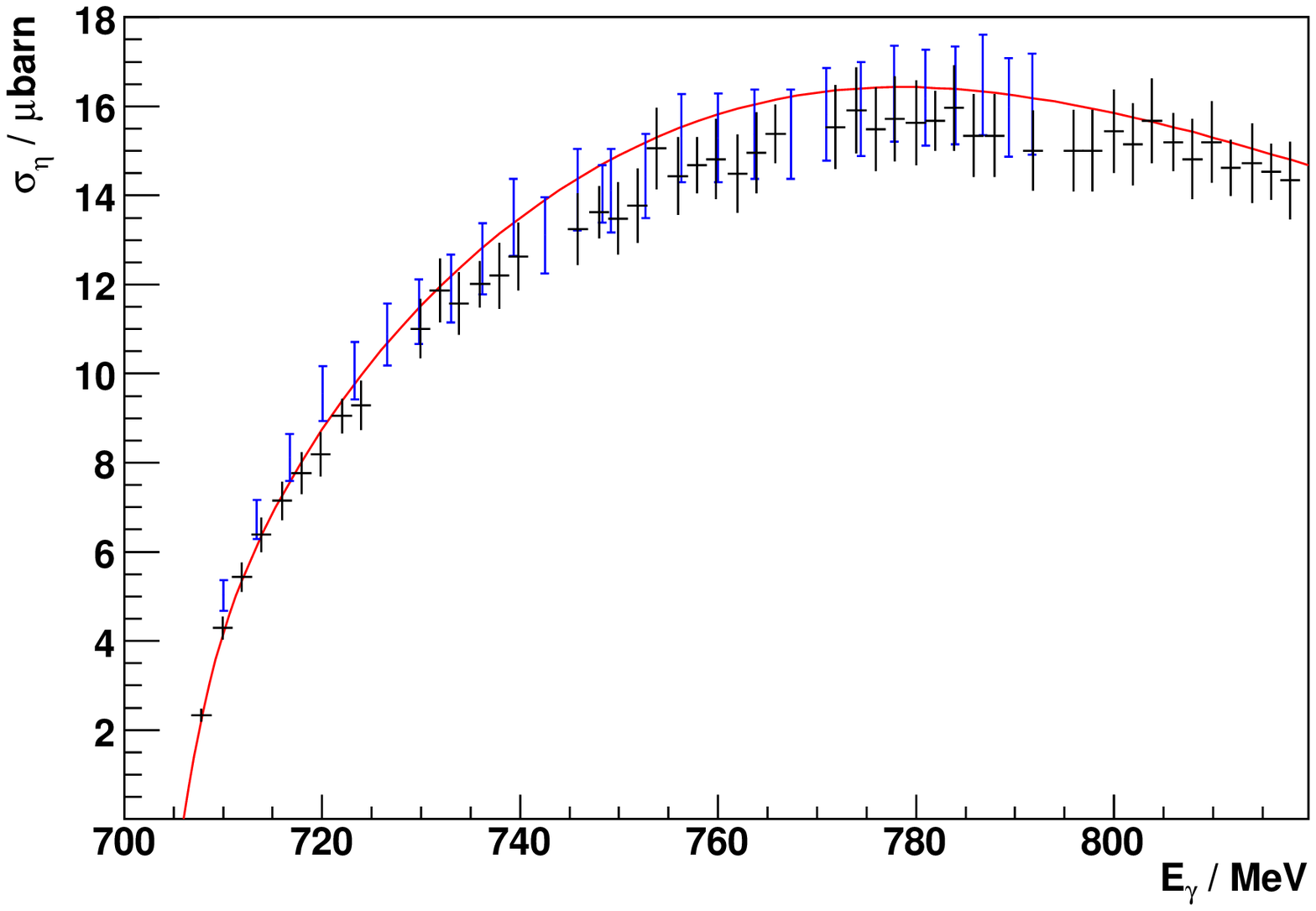}\includegraphics[scale=0.4]{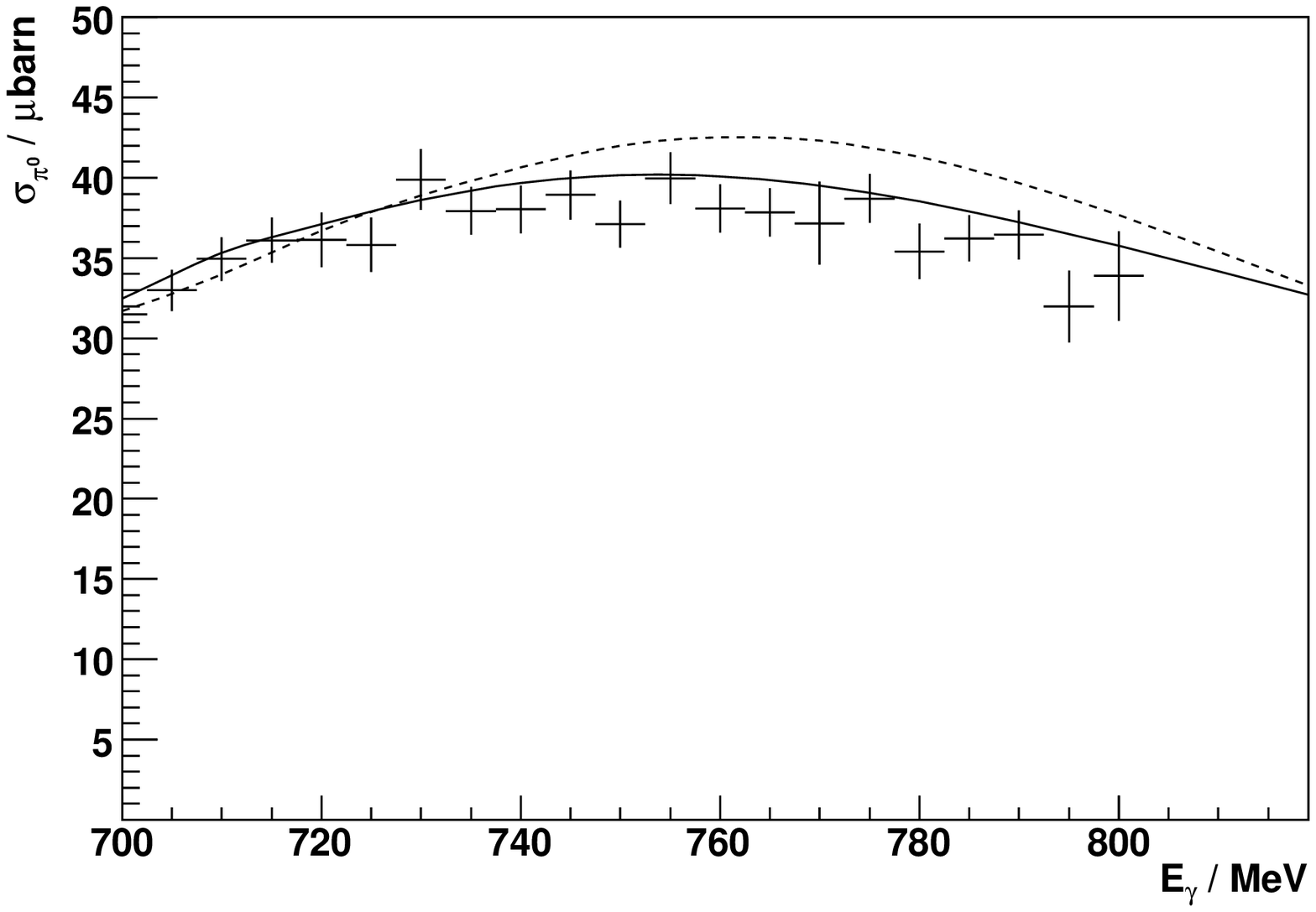}
		\caption{Total cross sections. \textit{Left}: $\eta$-photoproduction cross section compared to the measurement made with TAPS at MAMI \cite{Kru95} (points with only vertical error bars) and the Eta-MAID calculation \cite{Chi02} (solid line). \textit{Right}: $\pi^0$-photoproduction cross section compared to the MAID2007 \cite{Dre99,Dre07} (dashed line) and SAID \cite{Arn96} models (solid line).}
		\label{psWQ}
	\end{center}
\end{figure*}

In addition, good agreement between the experiment and the simulation had to be achieved in the distributions of the variables used as constraints in the kinematic fit. This is also illustrated in fig \ref{psFit}. Here the invariant mass of the six photons, calculated using the directly measured energies and momenta, is shown for the events selected by the kinematic fit. There is very good agreement between the two distributions, which have maxima at 547.7\,MeV and widths of $\sigma \approx 16$\,MeV. The missing mass of the measured photon 4-momenta is shown in the lower left plot. Both curves have a maximum at 936.7\,MeV and a width of $\sigma = 18.2$\,MeV, indicating that the chosen events satisfied this constraint. The $\pi^0$ mass spectrum is shown in the lower right plot, where the invariant masses of the three photon pairs from the fit with the highest CL are shown. The two distributions with peaks at 134.4\,MeV and widths of $\sigma \approx 7$\,MeV agree within 4\,\%.

As a further check of the analysis and the simulation, total cross sections for $\eta$- and $\pi^0$-photoproduction were determined and compared to previous experiments and model fits to experimental data. Figure \ref{psWQ} shows these cross sections for the energy range of 700 to 820\,MeV. The total $\eta$ cross section is compared to the measurement carried out with TAPS at MAMI \cite{Kru95} and the Eta-MAID model fit \cite{Chi02}, which is dominated by this TAPS experiment in the threshold region. It is clearly seen that the results obtained agree with both, the TAPS data and the Eta-MAID calculation. Figure \ref{psWQ} (right) shows the total $\pi^0$-photoproduction cross section as determined using an analysis similar to that described above. The only difference was that just two constraints, namely the invariant and the missing mass of two identified photons, were used. The $\pi^0$ cross section shows good agreement with the MAID2007 \cite{Dre99,Dre07} and SAID \cite{Arn96} models. The agreement is also very good in the region of the $\mathrm{\Delta}$ resonance, which is not shown here.

Although these comparisons indicate that the selected event sample was almost free of background, possible contaminating background reactions were considered. In the examined energy region the main back\-ground contribution comes from the direct $3\pi^0$-production through the reaction $\gamma \mathrm{p} \rightarrow 3\pi^0 \mathrm{p}$. To study the fraction of such background events in the selected $\eta$ sample $10^7$ $3\pi^0$ events were simulated and analysed, giving a total reconstruction efficiency $\varepsilon_{3\pi^0} \approx 5\,\%$. The background contribution was calculated using the estimate of the total cross section made in \cite{Jun05}, which resulted in $\sigma (\gamma \mathrm{p} \rightarrow 3\pi^0 \mathrm{p}) \approx 0.4\,\mu\mathrm{b}$ for photon beam energies $E_\gamma < 1100$\,MeV. It was assumed to be constant over the examined energy region. The contamination was then estimated to be
\begin{equation}
	\frac{N_{3\pi^0}}{N_{\eta \rightarrow 3\pi^0}} = \frac{\sigma (\gamma \mathrm{p} \rightarrow 3\pi^0 \mathrm{p})}{\bar{\sigma}_{\eta} \cdot \mathrm{BR}(\eta \rightarrow 3\pi^0)} \cdot \frac{\varepsilon_{3\pi^0}}{\varepsilon_{\eta \rightarrow 3\pi^0}} \approx 2\,\%,
\end{equation}
where $\varepsilon_{3\pi^0}$ and $\varepsilon_{\eta \rightarrow 3\pi^0}$ are the reconstruction efficiencies for the direct $3\pi^0$-production and the $\eta \rightarrow 3\pi^0$ decay, respectively, and BR($\eta \rightarrow 3\pi^0$) is the branching ratio for the given decay. The total $\eta$-photoproduction cross section was averaged over the observed photon beam energy range of 700\,MeV to 820\,MeV, resulting in $\bar{\sigma}_\eta \approx 14\,\mu\mathrm{b}$. The 2\,\% contribution is much smaller than the estimated statistical and systematic uncertainties (see section \ref{hdresults}), which were determined to be of the order of 5 to 10\,\% each. Therefore, it was neglected. Another background process, namely double $\pi^0$-production with two cluster split-offs, was also found to be negligible. Other background processes were kinematically not possible due to the restricted tagged photon energy range of $E_\gamma \leq 820$\,MeV.

\section{Results}\label{hdresults}

\subsection{Dalitz plot parameter}\label{hddalitz}

\begin{figure}
	\begin{center}
		\includegraphics[scale=0.4]{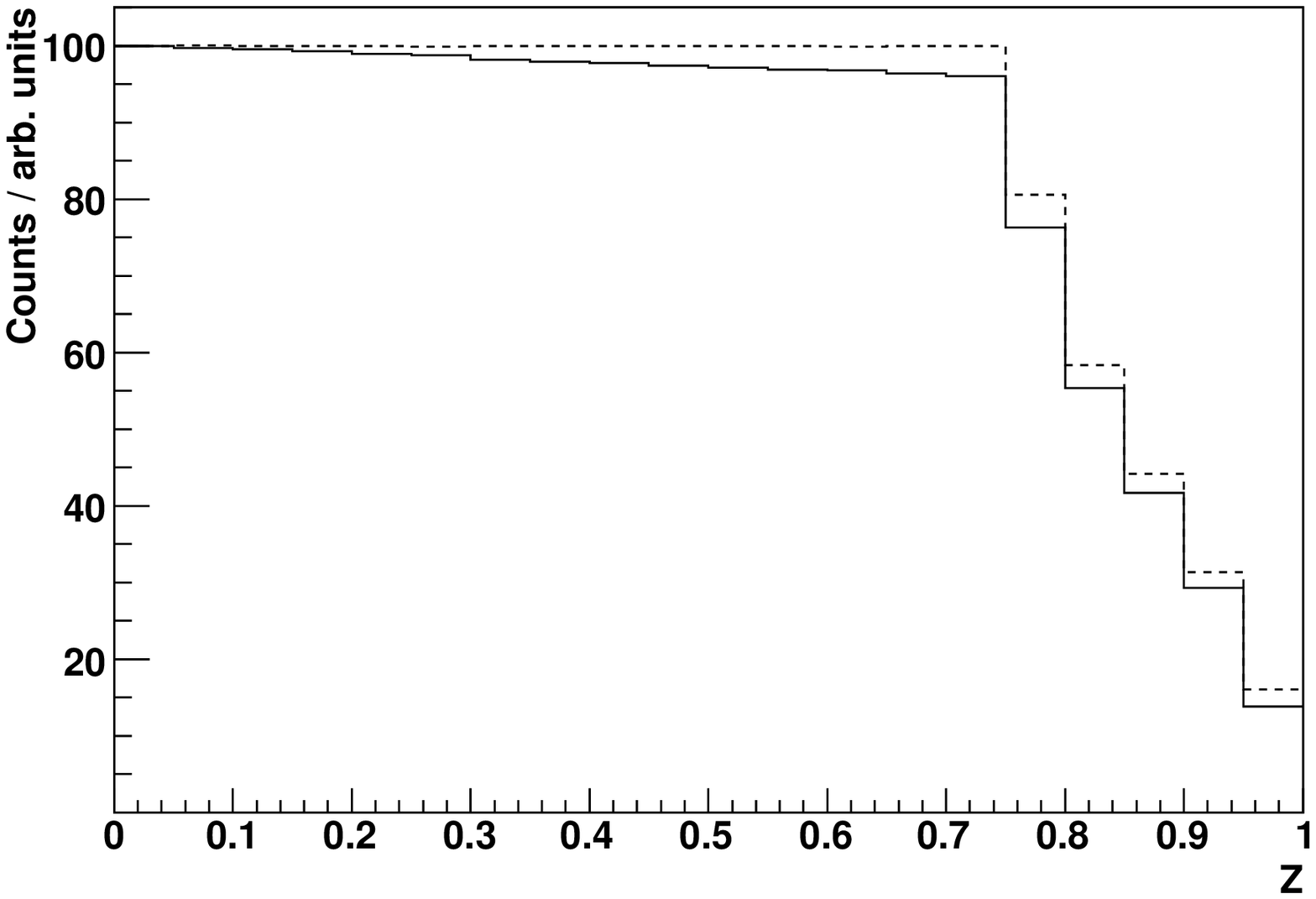}
		\includegraphics[scale=0.4]{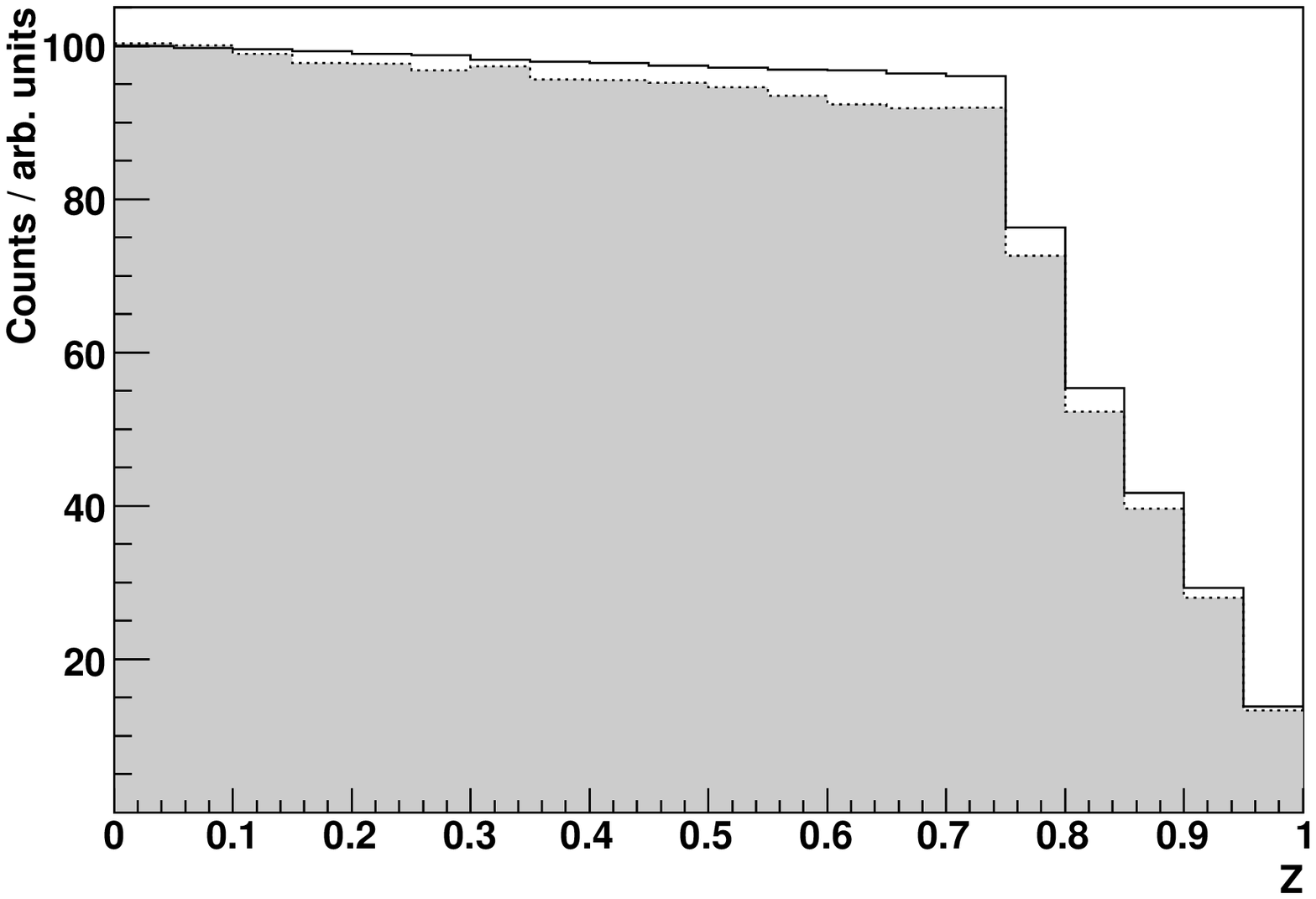}
		\caption{The $z$ distribution from the $\eta \rightarrow 3\pi^0$ decay events, selected by the kinematic fit. \textit{Top}: Comparison of the simulated distributions for complete detector acceptance and 100\,\% detection efficiency (dashed line) and for a realistic simulation (solid line). \textit{Bottom}: Comparison of the distributions reconstructed from the realistic simulation (same solid line as in the upper picture) and the measured data (grey shaded).}
		\label{psZ}
	\end{center}
\end{figure}

The Dalitz plot parameter, $\alpha$, for the $\eta \rightarrow 3\pi^0$ decay was determined by comparing the simulated with the measured $z$ distribution. The events were selected by kinematic fits testing the hypothesis $\gamma \mathrm{p} \rightarrow \eta \mathrm{p} \rightarrow 3\pi^0 \mathrm{p}$ at the 2\,\% CL. The simulation was based on pure phase-space distributions with $\alpha = 0$. The dashed line in fig. \ref{psZ} (top) shows the generated $z$ distribution for full detector acceptance and 100\,\% detection efficiency. The solid line illustrates that realistic acceptance and efficiency already introduce a slope to the simulated distribution, thus, affecting our result for the Dalitz plot parameter. From the distribution of the difference between reconstructed and initially generated $z$ values resolution for the variable $z$ was found to be $\sigma_z \approx 4.5\,\%$ and, thus, a bin width for the $z$ distribution of 0.05 was chosen.

\begin{figure}
	\begin{center}
		\includegraphics[scale=0.4]{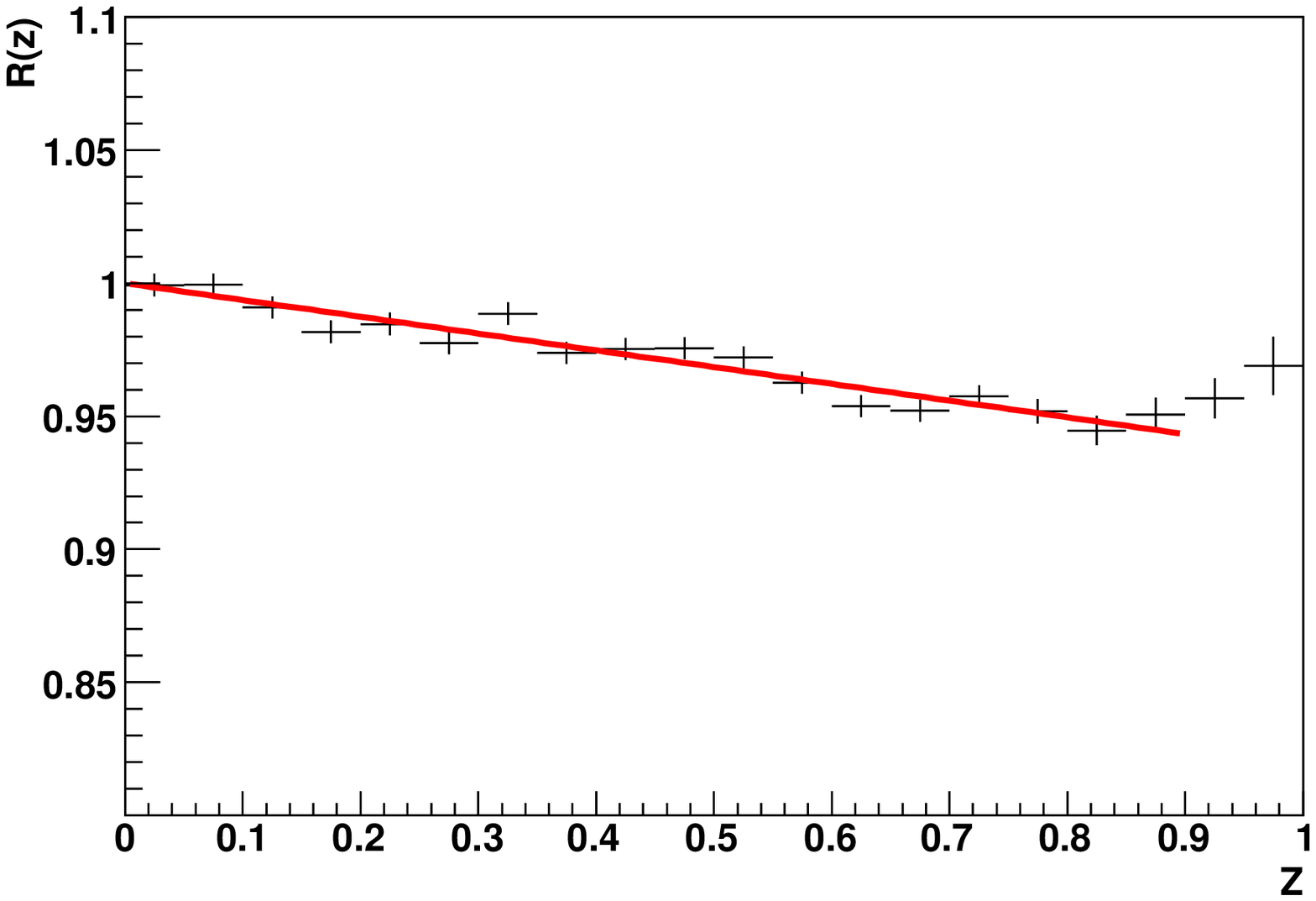}
		\includegraphics[scale=0.4]{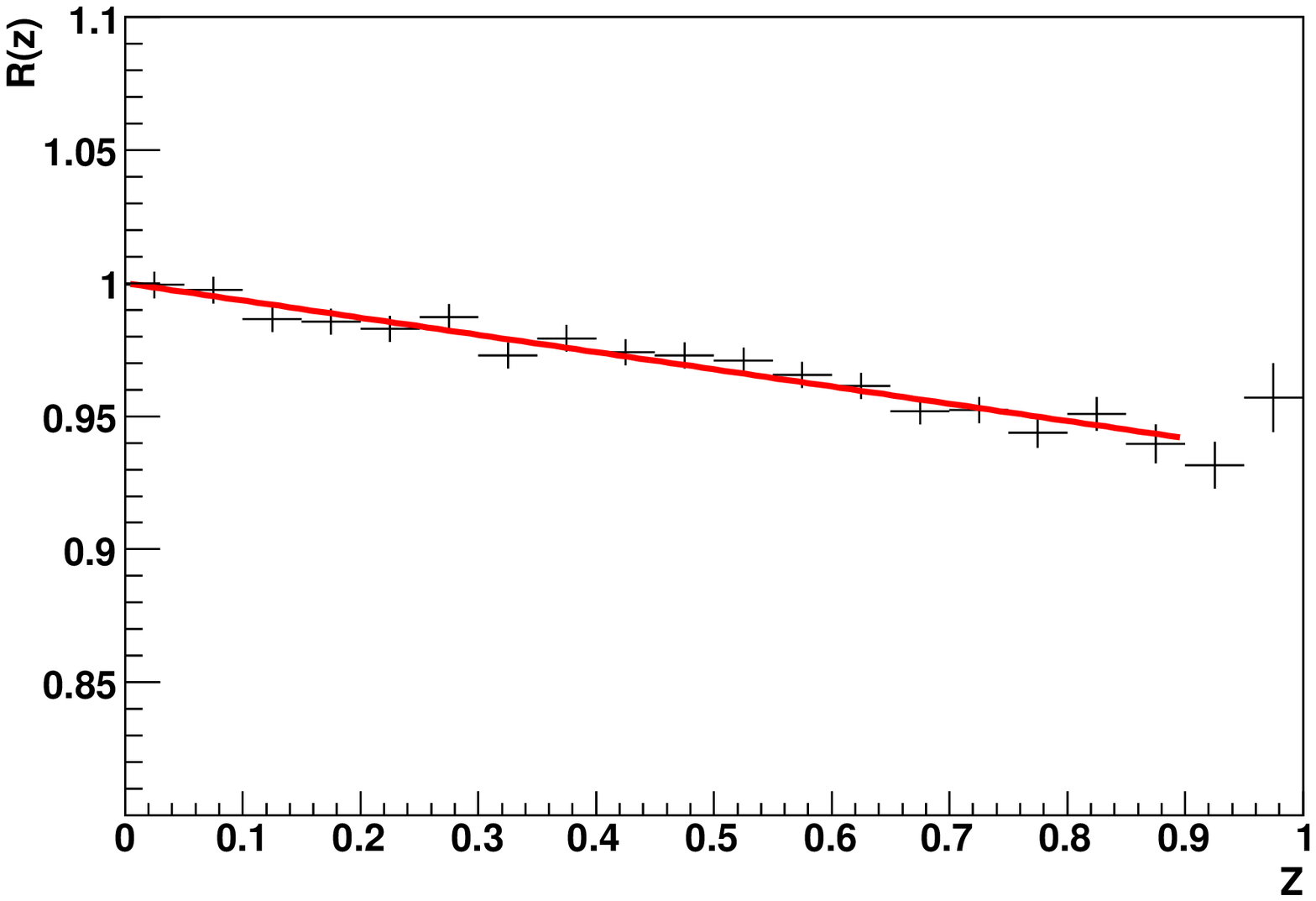}
		\caption{Ratio $R(z)$ of the simulated and the measured $z$ distributions for the two experiments. The Dalitz plot parameter, $\alpha$, was obtained directly from the fit of the function $c(1+2 \alpha z)$. In the figure, $R(z)$ has been scaled with $c$ so that the fitted lines have intercepts equal to 1 on the $R(z)$-axis. \textit{Top}: For the $\eta$ decays experiment. \textit{Bottom}: For the radiative $\pi^0$-production experiment.}
		\label{psSlope}
	\end{center}
\end{figure}

In order to determine the Dalitz plot parameter, the experimental $z$ distribution was divided by the simulated $z$ distribution. This ratio $R(z)$ was then fitted with the function $c(1+2 \alpha z)$, according to eq. \ref{eqAmpExpand}, using $c$ and $\alpha$ as free parameters in the fit. The deviation of $R(z)$ from pure phase space is then reflected in the value of $\alpha$. This deviation is illustrated in the lower picture of fig. \ref{psZ}. Here the simulated $z$ distribution was normalised with $c$. The ratio $R(z)$, also scaled with $c$, and the line fits for the two experiments are presented in fig. \ref{psSlope}. The upper limit of the fit region was fixed at $z=0.9$, because the last two bins had much poorer statistics and showed systematic deviations from a straight line. This effect occurred due to slightly different resolutions used in the kinematic fit for the experimental and the simulated data. These had to be applied to match the experimental and the simulated CL distributions. As the cut on the CL was the major restriction in the presented analysis, agreement in the CL distributions was desired. So, the last two bins were excluded by limiting the fit region to $z < 0.9$. The systematic effect of these two bins on the Dalitz plot parameter was examined in a test, fitting a line to the region $0 < z <1$ (see section \ref{hdsyst}).

The two experiments differed in the CB energy sum trigger threshold of $E_{\mathrm{thr}}^1 \approx 390$\,MeV and $E_{\mathrm{thr}}^2 \approx 60$\,MeV, and the tagged photon energies, 680\,MeV$\leq E_{\gamma}^1 \leq$820\,MeV compared with 200\,MeV$\leq~E_{\gamma}^2~\leq$820\,MeV. The results of the standard analysis described above for these two experiments, $\alpha_1$ and $\alpha_2$, are listed in the first line of table \ref{tbalpha}. The experimental statistics and the reduced $\chi^2$-values ($\chi^2$/ndf) derived from the fits to the $z$ distribution are also given. The listed statistical uncertainties were taken from the errors of the fits. The results from the two experiments agree very well.

\subsection{Systematic uncertainty}\label{hdsyst}

The systematic uncertainties were estimated in a series of tests, varying only one analysis parameter at a time. The results from all tests for both experiments are listed in table \ref{tbalpha}. For each test the difference compared to the standard analysis is indicated.

The first test demanded detection of exactly one proton in addition to the six photons as an event selection criterion ($6\gamma$+1p), requiring accepted events to have a detector cluster structure consistent with the $\gamma \mathrm{p} \rightarrow \eta \mathrm{p} \rightarrow 3\pi^0 \mathrm{p}$ reaction. The results of this test show, within one standard deviation, good agreement with the standard values for both experiments.

Then there was a series of tests where just the cut on the CL was altered between 1\,\% and 50\,\%. For the standard analysis it was set to 2\,\%. With the cut on the lower value of 1\,\% the effect of relaxing the acceptance criteria, and possibly including some events from background processes, was investigated. It showed small deviations compared to the standard analysis. Raising the cut to 5\,\%, 10\,\%, 20\,\% and 50\,\% led to a higher purity of the chosen event sample. Most results show slight differences compared to the standard value. Cuts on high confidence levels led to larger deviation. But, since at the same time the statistical significances decreased, the results are still within one standard deviation compared to the standard analysis.

\begin{table*}
	\caption{Results from tests concerning the systematic uncertainty of the extraction of the Dalitz plot parameter, $\alpha$, for the $\eta \rightarrow 3\pi^0$ decay. Given are abbreviations for the test conditions, which are explained in the main text, the test results with their statistics $N$ and the $\chi^2$/ndf of the straight line fit for both experiments.}
	\label{tbalpha}
	\begin{center}
	\begin{tabular}{cc|ccc|ccc}
		\hline\noalign{\smallskip}
		 & Test & $\alpha_1$ & $N_1$ / $10^3$ & $\chi^2$/ndf & $\alpha_2$ & $N_2$ / $10^3$ & $\chi^2$/ndf \\
		\noalign{\smallskip}\hline\noalign{\smallskip}
		1 & Standard & $-0.0315 \pm 0.0020$ & 1120 & 20/16 & $-0.0324 \pm 0.0024$ & 724 & 10/16 \\
		2 & 6$\gamma$+1p & $-0.0323 \pm 0.0033$ & 430 & 15/16 & $-0.0357 \pm 0.0035$ & 322 & 18/16 \\
		3 & CL=1 & $-0.0308 \pm 0.0020$ & 1177 & 20/16 & $-0.0331 \pm 0.0023$ & 759 & 9/16 \\
		4 & CL=5 & $-0.0327 \pm 0.0021$ & 1022 & 22/16 & $-0.0336 \pm 0.0025$ & 663 & 12/16 \\
		5 & CL=10 & $-0.0328 \pm 0.0022$ & 919 & 21/16 & $-0.0322 \pm 0.0026$ & 600 & 11/16 \\
		6 & CL=20 & $-0.0332 \pm 0.0024$ & 774 & 17/16 & $-0.0296 \pm 0.0028$ & 510 & 16/16 \\
		7 & CL=50 & $-0.0295 \pm 0.0031$ & 453 & 20/16 & $-0.0297 \pm 0.0036$ & 303 & 11/16 \\
		8 & $E_{\mathrm{thr}}=380$\,MeV & $-0.0323 \pm 0.0020$ & 1121 & 20/16 & $---$ & $---$ & $---$ \\
		9 & $E_{\mathrm{thr}}=388$\,MeV & $-0.0316 \pm 0.0020$ & 1120 & 20/16 & $---$ & $---$ & $---$ \\
		10 & $E_{\mathrm{thr}}=392$\,MeV & $-0.0313 \pm 0.0020$ & 1119 & 20/16 & $---$ & $---$ & $---$ \\
		11 & $E_{\mathrm{thr}}=400$\,MeV & $-0.0307 \pm 0.0020$ & 1117 & 21/16 & $---$ & $---$ & $---$ \\
		12 & no TAPS & $-0.0300 \pm 0.0024$ & 797 & 13/16 & $-0.0316 \pm 0.0028$ & 516 & 11/16 \\
		13 & no ID & $-0.0311 \pm 0.0023$ & 838 & 12/16 & $-0.0306 \pm 0.0027$ & 556 & 12/16 \\
		14 & $z\leq1$ & $-0.0295 \pm 0.0019$ & 1149 & 31/18 & $-0.0322 \pm 0.0022$ & 742 & 13/18 \\
		15 & $z<0.75$ & $-0.0319 \pm 0.0024$ & 1005 & 19/13 & $-0.0312 \pm 0.0029$ & 650 & 8/13 \\
		16 & $z<0.6$ & $-0.0284 \pm 0.0034$ & 812 & 13/10 & $-0.0277 \pm 0.0041$ & 524 & 7/10 \\
		\noalign{\smallskip}\hline
	\end{tabular}
	\end{center}
\end{table*}

The trigger threshold for the CB energy sum was determined for one experiment to be $E_{\mathrm{thr}}^1 \approx 390$\,MeV. Since this value could not be fixed exactly, in a series of tests it was varied around the standard value. But these tests were only performed for the experiment with the higher cut, because the threshold of the second experiment, roughly 60\,MeV, was so low that almost no $\eta \rightarrow 3\pi^0$ events were rejected (compare fig. \ref{psTrig}). The results of these tests showed good agreement with the standard value.

In another test TAPS clusters were ignored in the analysis (no TAPS), so that all six required photons had to be detected by the CB. This was a check, if the acceptance in forward directions was simulated precisely enough. Such a test was necessary, because a lot of inactive material in the exit region of the CB screened TAPS from particles emerging from the target. Both results are slightly lower than the standard values, but are within one $\sigma$ compared to them.

Possible systematic effects from misidentification of photons as charged particles were studied in an analysis (no ID) omitting the particle identification methods described above. All clusters, registered in the CB and TAPS, were accepted as photon candidates. Practically no deviations from the standard values were found.

As mentioned above, in a test the fit region was extended to $z=1.0$. With the fit over the full $z$ range the influence of the last two bins at high $z$ values was investigated. The results in table \ref{tbalpha} show no effect for the second, but a drop in $\alpha$ for the first experiment. Also the reduced $\chi^2$ indicates the the description of a straight line does not hold here for the full $z$ range. But within one standard deviation the test results are compatible with the standard values.

The upper limit of $z=0.75$ was chosen for a test to investigate possible influences of the region in the Dalitz plot with $z > 0.756$, where statistics significantly decrease (see fig. \ref{psZ}). In eq. \ref{eqz} $z$ is given in such a way that all points with $z \leq 0.756$ have the same probability for the pure phase space process. In \cite{Nis07} it is shown that at $z=0.756$ a cusp in $R(z)$ arises. The result of this test indicates just minor changes compared to the standard values.

As discussed in \cite{Nis07}, the ratio $R(z)$ exhibits two more cusps, one at $z=0.597$, the minimum value to reach the $\pi^+ \pi^-$ threshold in the Dalitz plot, and another at $z=0.882$, corresponding to the maximum value to touch this line. These cusps arise due to the $\pi^+ \pi^- \rightarrow \pi^0 \pi^0$ charge exchange-reaction, which also produces a cusp in the $\pi^0 \pi^0$ mass distribution of the $\eta \rightarrow 3\pi^0$ decay \cite{Dit08,Mei97,Mei98,Cab04,Cab05,Bel06,Bis07}. The cusp at higher $z$ in the ratio $R(z)$ was already excluded by the standard analysis. Reduction to $z=0.6$ makes a substantial decrease in the result for $\alpha$, but the statistical error is also larger. In \cite{Dit08} Ditsche, Kubis and Mei\ss ner calculated that, if the Dalitz plot parameter is obtained by a fit over the range $0 \leq z \leq 0.597$, a drop of 5\,\% in the absolute value of $\alpha$ should be visible, compared to a fit over the full $z$ range. The results from test 16 show a decrease of roughly 10\,\% and 15\,\% for the two experiments. But within the increased errors, the test results still agree with the standard values. So, no reliable statement on the influence of the cusps in $R(z)$ can be made and the differences in $\alpha$ were just taken into account in the determination of the systematic uncertainties.

The systematic uncertainties were then calculated for both experiments separately according to
\begin{equation}
	\Delta_{\mathrm{syst}}(\alpha_i) = \frac{\sum_k \left( \alpha^{\prime}_{ik} - \alpha_i \right) \cdot N_{ik}}{\sum_k N_{ik}}
        \label{eqsyserror}
\end{equation}
where $\alpha_i$ stands for the standard values of the two experiments. The $\alpha^{\prime}_{ik}$ are the results of the different tests listed in table \ref{tbalpha} with their statistics $N_{ik}$. Thus, the systematic uncertainties were calculated by the sum of the differences between the test results and the standard values, weighted with the statistics of the tests. Positive and negative deviations were handled separately in this process.

\subsection{Final results}\label{hdfinal}

As final results
\begin{equation}
	\alpha_1 = -0.0315 \pm 0.0020 ^{+0.0012}_{-0.0009}
\end{equation}
\begin{equation}
	\alpha_2 = -0.0324 \pm 0.0024 ^{+0.0016}_{-0.0014}
\end{equation}
were obtained. The result of the analysis presented in this paper was then calculated by the weighted mean of the two values, while the inverse variances were taken as weights. The statistical uncertainty was calculated from
\begin{equation}
	\sigma_{\mathrm{stat}}^2 = \frac{1}{\sum_i (1/\sigma_i^2)}.
\end{equation}
The systematic uncertainty was taken from the highest systematic uncertainty given above. As final result we quote
\begin{equation}
	\alpha = -0.032 \pm 0.002_{\mathrm{stat}} \pm 0.002_{\mathrm{syst}}
\end{equation}
This value is consistent with the current PDG average \cite{PDG08}, which is dominated by the Crystal Ball result measured at BNL, and agrees reasonably well with the results published by the KLOE \cite{Amb07}, CELSIUS/WASA \cite{Bas07} and WASA-at-COSY \cite{Ado08} collaborations.

After the upgrade of MAMI to 1.5\,GeV maximum electron energy \cite{Kai08}, a new experiment on the neutral decays of the $\eta$ meson was performed in the tagged photon energy range of 700\,MeV to 1400\,MeV. An independent analysis of this new experiment \cite{Pra08} obtained approximately $3 \cdot 10^6$ $\eta \rightarrow 3\pi^0$ events, giving a value for $\alpha$, which is in very good agreement with the result presented in this paper.

\section{Summary}\label{hdsummary}

The Dalitz plot parameter, $\alpha$, for the $\eta \rightarrow 3\pi^0$ decay has been measured with the Crystal Ball detector and TAPS at the electron accelerator facility MAMI-B in Mainz. The $\eta$ mesons were produced by bremsstrahlung photons emitted by the 883\,MeV electrons and detected by the Glasgow tagged photon spectrometer. The kinematic fitting analy\-sis selected $1.8 \cdot 10^6$ $\gamma \mathrm{p} \rightarrow \eta \mathrm{p} \rightarrow 3\pi^0 \mathrm{p}$ events and the result $\alpha = -0.032 \pm 0.002_{\mathrm{stat}} \pm 0.002_{\mathrm{syst}}$ is in good agreement with other high-precision experiments. A possible influence on the Dalitz plot parameter by the $\pi^+ \pi^- \rightarrow \pi^0 \pi^0$ contribution to the amplitude was found to be small, but is included in the systematic uncertainty.

\subsection*{Acknowledgements}

The authors wish to thank the accelerator group of MAMI for the precise and very stable beam conditions. This work was supported by the Deutsche Forschungsgemeinschaft (SFB 443, SFB/TR 16), the DFG-\-RFBR (Grant No. 05-02-04014), European Community-Research Infrastructure Activity under the FP6 ``Structuring the European Research Area programme" (HadronPhysics, Contract No. RII3-CT-2004-506078), the NSERC (Canada), Schweizerischer Nationalfonds, U.K. EPSRC, the U.S. DOE and U.S. NSF. We thank the undergraduate students of Mount Allison and George Washington Universities for their assistance.


\begin{thebibliography}{}

\bibitem{Sut66} D.G.~Sutherland, Phys. Lett. \textbf{23}, (1966) 384.
\bibitem{Bau96} R.~Baur, J.~Kambor and D.~Wyler, Nucl. Phys. \textbf{B460}, (1996) 127.
\bibitem{Dit08} C.~Ditsche, B.~Kubis and U.-G.~Mei{\ss}ner, to be published in Eur. Phys. J. \textbf{C}, (2008) [arXiv:0812.0344 [hep-ph]].
\bibitem{Osb70} H.~Osborn and D.J.~Wallace, Nucl. Phys. \textbf{B20}, (1970) 23.
\bibitem{Gas85} J.~Gasser and H.~Leutwyler, Nucl. Phys. \textbf{B250}, (1985) 539.
\bibitem{Bij07} J.~Bijnens and K.~Ghorbani, JHEP \textbf{0711}, (2007) 030.
\bibitem{Kam96} J.~Kambor, C.~Wiesendanger, D.~Wyler, Nucl. Phys. \textbf{B465}, (1996) 215.
\bibitem{Ani96} A.V.~Anisovich and H.~Leutwyler, Phys. Lett. \textbf{B375}, (1996) 335.
\bibitem{Bor05} B.~Borasoy and R.~Ni{\ss}ler, Eur. Phys. J. \textbf{A26}, (2005) 383.
\bibitem{PDG08} C.~Amsler \textit{et al.} (Particle Data Group), Phys. Lett. \textbf{B667}, (2008) 1.
\bibitem{Bij02} J.~Bijnens and J.~Gasser, Phys. Scripta \textbf{T99}, (2002) 34.
\bibitem{Khu60} N.N.~Khuri and S.B.~Treiman, Phys. Rev. \textbf{119}, (1960) 1115.
\bibitem{Eid04} S.~Eidelman \textit{et al.} [Particle Data Group], Phys. Lett. \textbf{B592}, (2004) 1.
\bibitem{Tip01} W.B.~Tippens \textit{et al.}, Phys. Rev. Lett. \textbf{87}, (2001) 192001.
\bibitem{Amb07} F.~Ambrosino \textit{et al.}, arXiv:0707.4137 [hep-ex].
\bibitem{Bag70} C.~Baglin \textit{et al.}, Nucl. Phys. \textbf{B22}, (1970) 66.
\bibitem{Ald84} D.~Alde \textit{et al.}, Z. Phys. \textbf{C25}, (1984) 225.
\bibitem{Abe98} A.~Abele \textit{et al.}, Phys. Lett. \textbf{B417}, (1998) 193.
\bibitem{Ach01} M.N.~Achasov \textit{et al.}, JETP Lett. \textbf{73}, (2001) 451.
\bibitem{Bas07} M.~Bashkanov \textit{et al.}, Phys. Rev. \textbf{C76}, (2007) 048201.
\bibitem{Ado08} C.~Adolph \textit{et al.}, arXiv:0811:2763 [nucl-ex].
\bibitem{Cap05} T.~Capussela \textit{et al.}, Acta Phys. Slov. \textbf{56}, (2005) 341.
\bibitem{Her83} H.~Herminghaus \textit{et al.}, IEEE Trans. Nucl. Sci. \textbf{30}, (1983) 3274.
\bibitem{Wal90} Th.~Walcher, Prog. Part. Nucl. Phys. \textbf{24}, (1990) 189.
\bibitem{Unv08} M.~Unverzagt, PhD thesis, Rheinische Friedrich-Wilhelms-Uni\-versit\"at Bonn, Germany, 2007.
\bibitem{Ant91} I.~Anthony, J.D.~Kellie, S.J.~Hall, G.J.~Miller and J.~Ahrens, Nucl. Instrum. Meth. \textbf{A301}, (1991) 230.
\bibitem{Hal96} S.J.~Hall, G.J.~Miller, R.~Beck and P.~Jennewein, Nucl. Instrum. Meth. \textbf{A368}, (1996) 698.
\bibitem{Ore82} M.~Oreglia \textit{et al.}, Phys. Rev. \textbf{D25}, (1982) 2259.
\bibitem{Sta01} A.~Starostin \textit{et al.}, Phys. Rev. \textbf{C64}, (2001) 055205.
\bibitem{Wat04} D.~Watts, in \textit{Calorimetry in Particle Physics: Proceedings of the 11th International Conference, Perugia, Italy, 2004}, edited by C.~Cecchi, P.~Lubrano and M.~Pepe (World Scientific, Singapore, 2005), p. 560.
\bibitem{Nov91} R.~Novotny, IEEE Trans. Nucl. Sci. \textbf{38}, (1991) 379.
\bibitem{Gab94} A.R.~Gabler \textit{et al.}, Nucl. Instrum. Meth. \textbf{A346}, (1994) 168.
\bibitem{Ave91} P.~Avery, \textit{http://www.phys.ufl.edu/\begin{scriptsize}$\sim$\end{scriptsize}avery/fitting.html}.
\bibitem{Kru95} B.~Krusche \textit{et al.}, Phys. Rev. Lett \textbf{75}, (1995) 3023.
\bibitem{Chi02} W.-T.~Chiang, S.N.~Yang, L.~Tiator, M.~Vanderhaegen, D.~Drechsel, Phys. Rev. \textbf{C68}, (2003) 045202.
\bibitem{Dre99} D.~Drechsel, O.~Hanstein, S.S.~Kamalov, L.~Tiator, Nucl. Phys. \textbf{A645}, (1999) 145.
\bibitem{Dre07} D.~Drechsel, S.S.~Kamalov, L.~Tiator, Eur. Phys. J. \textbf{A34}, (2007) 69.
\bibitem{Arn96} R.A.~Arndt \textit{et al.}, Phys. Rev \textbf{C53}, (1996) 430.
\bibitem{Jun05} J.~Junkersfeld, PhD thesis, Rheinische Friedrich-Wilhelms-Universit\"at Bonn, Germany, 2005.
\bibitem{Nis07} R.~Ni\ss ler, PhD thesis, Rheinische Friedrich-Wilhelms-Uni\-versit\"at Bonn, Germany, 2007.
\bibitem{Mei97} U.-G.~Mei\ss ner, G.~M\"uller and S.~Steininger, Phys. Lett. \textbf{B406}, (1997) 154 [Erratum-ibid. \textbf{B407}(1997) 454].
\bibitem{Mei98} U.-G.~Mei\ss ner, Nucl. Phys. \textbf{A629}, (1998) 72C.
\bibitem{Cab04} N.~Cabibbo, Phys. Rev. Lett. \textbf{93}, (2004) 121801.
\bibitem{Cab05} N.~Cabibbo, G.~Isidori, JHEP \textbf{0503}, (2005) 021.
\bibitem{Bel06} J.~Belina, Diploma thesis, Universit\"at Bern, Switzerland, 2006.
\bibitem{Bis07} M.~Bissegger, A.~Fuhrer, J.~Gasser, B.~Kubis and A.~Rusetsky, Phys. Lett. \textbf{B659}, (2008) 576.
\bibitem{Kai08} K.-H.~Kaiser \textit{et al.}, Nucl. Instrum. Meth. \textbf{A593}, (2008) 159.
\bibitem{Pra08} S.~Prakhov \textit{et al.}, submitted to Phys. Rev. \textbf{C}, (2008) [arXiv:0812.1999v2 [hep-ex]].

\end{thebibliography}
\end{document}